\documentclass[%
 reprint,
 amsmath,amssymb,
 aps,
]{revtex4-2}

\usepackage{graphicx}
\usepackage{dcolumn}
\usepackage{bm}
\usepackage{hyperref}
\usepackage[mathlines]{lineno}
\usepackage{array}
\usepackage{amsmath}

\usepackage{amssymb,amsmath}
\usepackage{graphicx}
\usepackage[dvipsnames]{xcolor}
\usepackage{physics}
\usepackage{bm}
\usepackage{bbm}
\usepackage[tight]{subfigure}
\usepackage[export]{adjustbox}
\usepackage{enumerate}
\usepackage{appendix}
\usepackage{fancyhdr}
\usepackage{siunitx}
\usepackage{multirow}
\usepackage{rotating,booktabs}
\usepackage[verbose]{placeins}
\usepackage{dsfont}
\usepackage{mathrsfs}
\usepackage{makecell}
\newcolumntype{T}[1]{S[table-format=#1]}

\newcolumntype{F}[1]{%
    >{\raggedright\arraybackslash\hspace{0pt}}p{#1}}%

\begin{document}

\preprint{APS/123-QED}

\title{Dependence of the gas phase high harmonic generation process on the driving laser--field duration and intensity}

\author{Bal\'azs Nagyill\'es$^{1,2}$,  Szabolcs T\'oth$^{1}$, Prabhash Prasannan Geetha$^{1}$, Paraskevas Tzallas$^{1,3,4}$, Zsolt Diveki$^{1}$}
 \email{zsolt.diveki@eli-alps.hu}

\affiliation{%
$^1$ELI ALPS, ELI-HU Non-Profit Ltd., Wolfgang Sandner utca 3., Szeged 6728, Hungary\\
$^2$Institute of Physics, University of Szeged, D\'om t\'er 9, H-6720 Szeged, Hungary\\
$^3$Foundation for Research and Technology - Hellas, Institute of Electronic Structure \& Laser, PO Box 1527, GR71110 Heraklion (Crete), Greece\\
$^4$Center for Quantum Science \& Technologies (FORTH-QuTech), GR-70013 Heraklion (Crete), Greece
}%

\date{\today}

\begin{abstract}

Ever since the advent of high-order harmonic generation, one of the main goals has been to maximize the high harmonic yield. This is due to the wide range of applications in multidisciplinary research fields, including nonlinear XUV optics and ultrafast science. Nowadays, intense laser–atom interactions are one of the primary sources of high-order harmonic generation, emitting radiation in the extreme ultraviolet (XUV) range. Although the scaling laws for XUV photon numbers have been extensively studied in the past, their dependence on the duration of the driving laser pulse has remained largely unexplored experimentally. This is because, in each of these studies, the XUV photon yield was optimized according to the specific characteristics of the lasers used and the corresponding XUV beamlines. In other words, there have been no systematic measurements on the dependence of XUV yield on the duration of the driving laser pulse. Here, by taking advantage of the SYLOS laser system at ELI–ALPS, we are able to experimentally investigate this long-standing question. We found that for driving laser field intensities below the saturation threshold of harmonic generation, the harmonic yield depends linearly on the pulse duration. However, for laser intensities near to the harmonic yield saturation, the harmonic yield is inversely proportional to the pulse duration.

\end{abstract}

\maketitle

\section{Introduction}
\begin{figure}[htbp]
    \centering
    \includegraphics[width=0.48\textwidth]{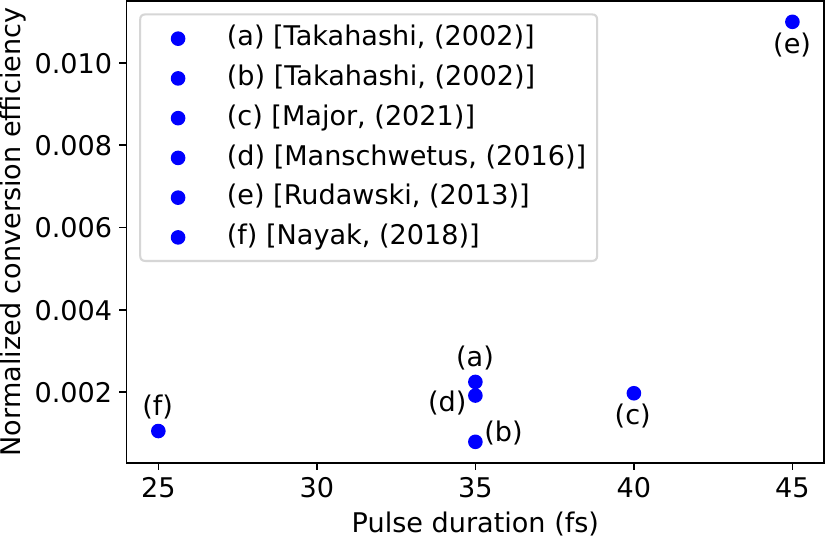}
    \caption{Reported normalized conversion efficiencies (N--CE) as a function of the pulse duration. The points have been obtained by Table. I}
    \label{fig:normalized_conversion}
\end{figure}

High-harmonic generation (HHG) is a nonlinear optical process that has revolutionized the field of ultrafast science by enabling the generation and characterization of attosecond pulses directly from tabletop laser systems \cite{RevMod_lhullier,RevMod_krausz,RevMod_agostini}.  These pulses, produced via HHG, serve as powerful tools for conducting time-resolved spectroscopy to study ultrafast phenomena, like probing electron dynamics within atoms, molecules \cite{Manschwetus2016, Orfanos2022}, and condensed matter \cite{Lucchini2016, Jordan2020,Nayak_2025} with unprecedented temporal resolution. However, the relatively low XUV photon flux delivered by these sources challenges their applicability in wide range of applications especially those related with nonlinear XUV optics and ultrafast science. Therefore, it is paramount to understand what are the essential parameters to optimize the HHG yield. There are numerous theoretical \cite{Weissenbilder2022, Heyl2016, Lewenstein1994, LHuillier1991,Brabec2000, Gaarde2008, Constant1999} and experimental \cite{Appi2023,Kazamias2003, Nefedova2018, Hergott2002,Tamaki2000, Delfin1999,Shiner2009} studies investigating this issue. 
Different laboratories have reported a range of harmonic fluxes making direct comparison challenging due to the different generating conditions they used. Table \ref{tab:nce} shows some representative published results concerning the measured photon number of the XUV radiation produced by the interaction of Argon (Ar) gas with intense IR laser pulses. We note that the XUV photon numbers placed on the table have been obtained from the published works taking into account the XUV transposition as has been reported in these papers. As conversion efficiency (CE) and normalize CE (N--CE), we define $\text{CE}=E_{\text{XUV}}/E_{\text{IR}}$ and $\text{N-CE}=\text{CE}/A$, receptively. $E_{\text{IR,XUV}}$ is the energy of the driving IR and XUV fields, $A$ is the focal spot area in the HHG medium. The diversity of the results is clearly shown in Fig. \ref{fig:normalized_conversion} which shows the dependence of the N--CE on the duration of the driving laser field used in these works.

\begin{table}\label{tab:nce}

\begin{tabular}{c c c c c c c}

\hline
\bfseries\makecell{E$_{\text{XUV}}$ \\ (nJ)}&
\bfseries\makecell{E$_{\text{IR}}$ \\ (mJ)}&
\bfseries\makecell{A \\ (cm$^2$)}&

\bfseries\makecell{$\tau$ \\ (fs)}&
\textbf{CE}&
\bfseries\makecell{N-CE \\ (cm$^{-2}$eV$^{-1}$)}&
\textbf{Ref.}\\ 
\hline

    $700$ & $20$ & $2.84E-04$ & $35$ & $3.50E-05$ & $2.24E-03$ & \cite{Takahashi2002} \\  
    
    $300$ & $20$ & $3.46E-04$ & $35$ & $1.50E-05$ & $7.87E-04$ & \cite{Takahashi2002} \\  
    
    $600$ & $16$ & $3.14E-04$ & $40$ & $3.75E-05$ & $1.97E-03$ & \cite{Major2021Opt} \\ 
     
    $800$ & $80$ & $9.50E-05$ & $35$ & $1.00E-05$ & $1.91E-03$ &  \cite{Manschwetus2016} \\  
    
    $1150$ & $20$ & $9.50E-05$ & $45$ & $5.75E-05$ & $1.10E-02$ & \cite{Rudawski2013} \\  
    
    $10000$ & $45$ & $3.85E-03$ & $25$ & $2E-04$ &  $1.05E-03$ & \cite{Nayak2018} \\ 

\hline

\end{tabular}

\caption{Reported photon energy, conversion efficiency (CE) and normalized to the focal spot area $A$ conversion efficiency (N--CE) of the plateau in Al filter transmission window (55 eV bandwidth) generated by the interaction of Argon gas with IR fs pulses of intensities in the range of $10^{14}$ to $3\times 10^{14}$ W/cm$^2$. It is noted that all the values in the table are approximate. For details we provide the corresponding references.}

\label{tab:list}
\end{table}
Generally speaking there are two main factors influencing the harmonic yield exiting the HHG medium: the single atom response \cite{Krause1992,Corkum1993, Lewenstein1994} and the phase matching in the macroscopic medium \cite{Brabec2000, Gaarde2008, Constant1999,Weissenbilder2022}. The former depends on the laser parameters (like beam energy, wavefront, pulse duration and wavelength) and the target medium, while the latter is influenced by the focusing geometry, the pressure and the medium length. Although laser pulse duration is one of the most crucial parameter and has been studied theoretically \cite{Brabec2000,Tempea2000,Shi2017,Holkundkar2023}, there is a significant lack of experimental research directly examining its impact on HHG yield, especially using sub 30\,fs pulses. 
Previous studies, such as those by Christov et al. \cite{Christov1996,Christov1997}, have demonstrated that laser pulse duration plays a critical role in extending the harmonic cutoff and enhancing HHG efficiency. Additionally, Tempea and Brabec \cite{Tempea2000} highlighted the existence of an optimal pulse duration for maximizing harmonic output in their numerical simulations. Very recently, Westerberg et al. \cite{Westerberg2025} arrived to similar conclusion from their experimental data in the 40-180\,fs range. However, a comprehensive experimental investigation into how varying pulse duration influences the harmonic yield, particularly in terms of achieving the highest photon flux including few cycle laser pulses, remains largely unexplored. This is due to the demanding requirements regarding laser specifications.

This study aims to partially fill this gap by systematically examining the influence of laser pulse duration on HHG yield below 30\,fs. Through a combination of experimental measurements and theoretical simulations, we explore how different pulse durations affect both the single-atom response and phase-matching conditions, with the goal of identifying the optimal conditions for maximizing photon flux. 
We experimentally demonstrate that for driving field intensities below the saturation threshold of harmonic generation process, the harmonic yield depends linearly on the pulse duration. The experiment has been conducted by changing the duration of the driving laser while all the other parameters are kept constant. The laser pulse width was kept transform limited to avoid any complications coming from chirping the laser pulse \cite{Csizmadia2021}. The results found to be in agreement with the theoretical model used to describe the HHG process. Using this model we have simulated the dependence of the harmonic yield on the driving field pulse duration for driving intensities close the to the saturation of the harmonic yield--conditions that are typically used to maximize the harmonic yield. In this case, we have found that the harmonic yield is inversely proportional to the pulse duration.
This research not only advances the fundamental understanding of pulse duration effects in HHG but also provides practical guidelines for enhancing the performance of tabletop attosecond sources.

\section{Simulation method}

To analyze the experimental results in high-harmonic generation (HHG), we employ advanced numerical simulations. These simulations account for macroscopic propagation effects—including plasma formation, absorption, and refraction—that are known to play critical roles in determining the efficiency of phase matching and, consequently, the total harmonic yield. To accurately model these effects, we perform comprehensive three-dimensional (3D) non-adiabatic macroscopic simulations based on models described in Refs.~\cite{Tosa2009,Priori2000,Major2019}.

The simulation proceeds through three self-consistent computational stages. The first step involves calculating the propagation of the linearly polarized driving laser field \( E(\mathbf{r}_L,t) \) within the nonlinear medium. This is governed by a modified scalar wave equation of the form

\begin{equation}
\nabla^2 E(\mathbf{r}_L,t) - \frac{1}{c^2} \frac{\partial^2 E(\mathbf{r}_L,t)}{\partial t^2} = \frac{\omega_0^2}{c^2} \left[ 1 - n_\mathrm{eff}^2(\mathbf{r}_L,t) \right] E(\mathbf{r}_L,t),
\label{eq:laser_wave_equation}
\end{equation}

where \( c \) is the speed of light in vacuum, \( \omega_0 \) is the central angular frequency of the laser, and \( \mathbf{r}_L \) denotes the spatial coordinate system aligned along the laser propagation axis. The effective refractive index \( n_\mathrm{eff}(\mathbf{r}_L,t) \) includes linear and nonlinear contributions and is given by

\begin{equation}
n_\mathrm{eff}(\mathbf{r}_L,t) = n + \bar{n}_2 I(\mathbf{r}_L,t) - \frac{\omega_p^2(\mathbf{r}_L,t)}{2 \omega_0^2},
\end{equation}

where \( I(\mathbf{r}_L,t) = \frac{1}{2} \epsilon_0 c |\tilde{E}(\mathbf{r}_L,t)|^2 \) is the laser intensity envelope calculated from the complex electric field \( \tilde{E}(\mathbf{r}_L,t) \), and \( \omega_p(\mathbf{r}_L,t) = \sqrt{n_e(\mathbf{r}_L,t) e^2 / (m \epsilon_0)} \) is the plasma frequency, determined by the local electron density \( n_e \). The constants \( e \), \( m \), and \( \epsilon_0 \) denote the electron charge, mass, and vacuum permittivity, respectively. This formulation allows incorporation of Kerr self-focusing, dispersion, and plasma defocusing effects, with absorption losses modeled using ionization rates as described in Ref.~\cite{Geissler1999}.

Assuming cylindrical symmetry about the laser propagation axis and employing the paraxial and slowly-evolving envelope approximations, the time-dependence in Eq.~\eqref{eq:laser_wave_equation} is eliminated via Fourier transform. In a co-moving frame, the wave equation becomes

\begin{multline}\label{eq:fourier_wave_equation}
\left( \frac{\partial^2}{\partial r_L^2} + \frac{1}{r_L} \frac{\partial}{\partial r_L} \right) E(r_L,z_L,\omega) - \frac{2 i \omega}{c} \frac{\partial E(r_L,z_L,\omega)}{\partial z_L} \\
= \frac{\omega^2}{c^2} \mathscr{F} \left[ \left( 1 - n_\mathrm{eff}^2(r_L,z_L,t) \right) E(r_L,z_L,t) \right],
\end{multline}

where \( \mathscr{F} \) denotes the Fourier transform in time. This equation is solved using a Crank–Nicolson finite-difference method, and the initial laser field distribution at the input plane is defined via ABCD-Hankel transform \cite{Ibnchaikh2001,Major2018AO}.

In the second stage, the microscopic dipole response is computed at each spatial grid point \((r_L,z_L)\) using the Lewenstein model \cite{Lewenstein1994}. The dipole acceleration \( D(t) \) is obtained from the local laser field and used to construct the nonlinear polarization response. Ground state depletion is incorporated by calculating the macroscopic polarization \( P_\mathrm{nl}(t) \) as

\begin{equation}
P_\mathrm{nl}(t) = n_a D(t) \exp \left[ - \int_{-\infty}^t w(t') \, dt' \right],
\end{equation}

where \( n_a \) is the local atomic number density and \( w(t) \) is the ionization rate computed using the hybrid antisymmetrized coupled channels (haCC) method \cite{Majety2015}, which shows strong agreement with the Ammosov–Delone–Krainov (ADK) model \cite{ADK1986}.

In the final step, the harmonic field \( E_h(\mathbf{r}_L,t) \) generated by the nonlinear polarization is propagated by solving the inhomogeneous wave equation

\begin{equation}
\nabla^2 E_h(\mathbf{r}_L,t) - \frac{1}{c^2} \frac{\partial^2 E_h(\mathbf{r}_L,t)}{\partial t^2} = \mu_0 \frac{\partial^2 P_\mathrm{nl}(t)}{\partial t^2},
\label{eq:harmonic_wave_equation}
\end{equation}

where \( \mu_0 \) is the vacuum permeability. The solution procedure is analogous to that used for the fundamental field, but here the source term is explicitly known. Dispersion and absorption in the extreme ultraviolet (XUV) regime are included through a complex refractive index model, with parameters derived from tabulated atomic scattering factors \cite{Henke1993}. The real part modifies the phase velocity, while the imaginary part accounts for absorption of the harmonic field during propagation.

\section{Experimental setup}
\begin{figure}[htbp]
    \centering
    \includegraphics[width=0.48\textwidth]{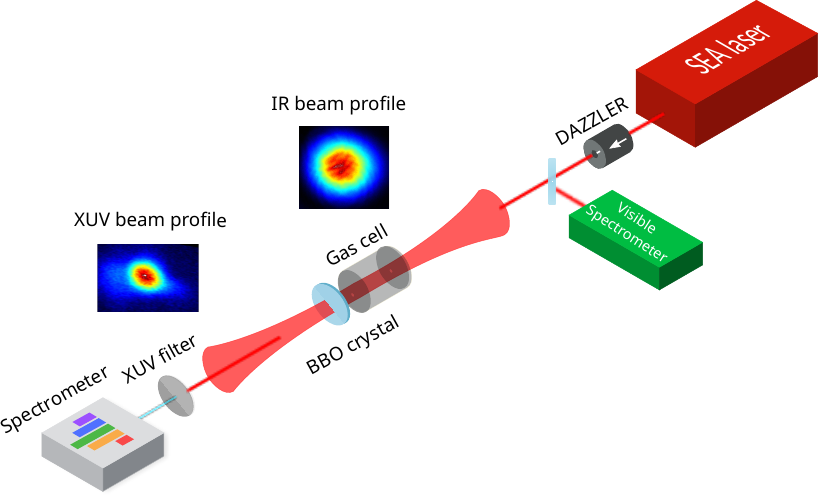}
    \caption{Schematic drawing of the experimenatal setup. The laser pulse is focused into a gas cell to perform HHG. Subsequently, a metallic XUV filter removes the IR beam, while the XUV is guided into an imaging XUV spectrometer. The spectral shaping of the laser was achieved with a DAZZLER and was detected with a visible spectrometer after an insertable mirror in the main beampath. The laser pulse duration was measured with chirp-scan using a BBO crystal after the target gas cell.}
    \label{fig:beamline}
\end{figure}

\begin{figure}[htbp]
    \centering
    \includegraphics[width=0.48\textwidth]{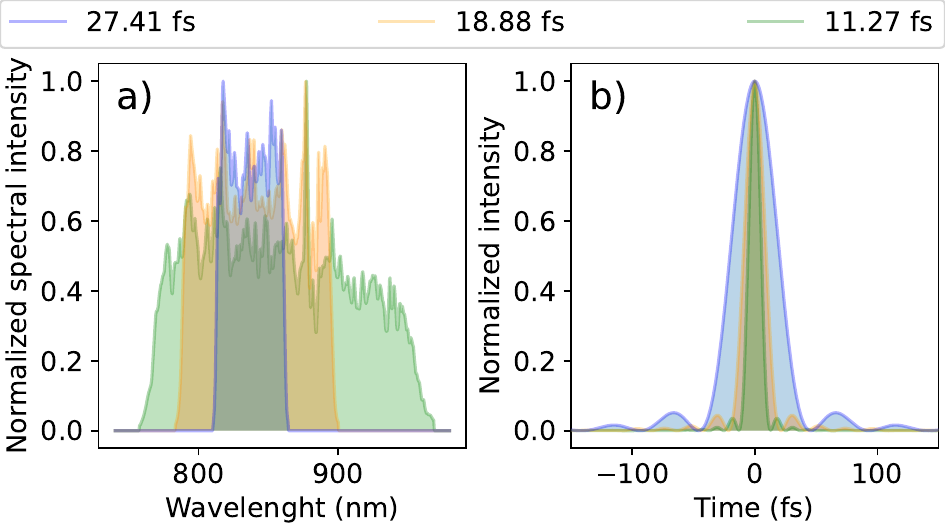}
    \caption{Three representative measured laser spectra (a) by symmetrically cutting the laser bandwidth and the corresponding transform limited pulse durations (b). }
    \label{fig:laser_spectra}
\end{figure}

The experiments were conducted at the SYLOS Long beamline of Extreme Light Infrastructure Attosecond Light Pulse Source (ELI-ALPS) facility \cite{Kuhn2017,Shirozhan2024}. The driving laser source was the SEA (SYLOS Experimental Attosecond) system \cite{Budriunas2017}, capable of delivering 35 mJ pulses with a central wavelength of 850 nm, a temporal duration of 12 fs (transform-limited), and a repetition rate of 10 Hz. These pulses were used to drive high-harmonic generation under loose focusing conditions, as schematically depicted in Fig. \ref{fig:beamline}. The diameter of the IR laser beam entering the XUV beam line was  $6$\,cm. In order to minimize spatiotemporal effects that may affect the laser beam properties in the HHG medium we have employed a deformable mirror with a set focal length of 18\,m. The interaction took place in a 300 mm long static gas cell filled with Ar gas of $1$\,mbar pressure. The focus position was placed close to the position of the Ar cell. The confocal parameter was comparable with the cell size i.e. $\sim 30$\,cm. The measured XUV beam profile was close to Gaussian \textbf{(Fig. \ref{fig:beamline})}. After the cell, an Al foil was implemented to remove the IR pulse. The remaining XUV radiation was guided into an imaging XUV spectrometer.
In order to control the spectral bandwidth and temporal duration of the infrared (IR) pulses, an acousto-optic programmable dispersive filter (AOPDF), specifically a DAZZLER device, was employed. This allowed precise shaping of the spectral amplitude and phase of the driving pulses, thereby enabling the preparation of pulses with tunable chirp and controlled transform-limited durations. To verify the effective pulse duration after spectral shaping, the chirp scan technique was used, which is based on the optimization of the second harmonic generated in a BBO crystal \cite{Loriot2013}. The resulting pulse durations for a selection of bandwidth settings are presented in Fig.~\ref{fig:laser_spectra}. The described transform limited pulses had a pulse duration ranging from 11.27\,fs to 40.69\,fs. 

It is important to emphasize that truncating the spectral bandwidth using the DAZZLER inherently reduces the pulse energy, as the optical energy contained within the filtered-out spectral wings is lost. This trade-off between pulse duration and pulse energy is a critical parameter in optimizing HHG efficiency and was carefully accounted for in both the experimental procedure and the accompanying macroscopic simulations. The systematic study of the effect of laser pulse duration on the harmonic yield has not been ever studied on the presented range of pulse duration because it requires high power ultrafast lasers like the ones at ELI-ALPS.

\section{Dependence of the harmonic yield on the IR pulse duration}

To investigate the effect of laser pulse duration on the harmonic yield, one should find the threshold laser intensity ($I_{\text{IR}}^{(th)}$), from which above the harmonic yield is not growing, for each transform limited laser pulse duration. This is a very demanding criteria in our case, since the transform limited duration is achieved by cutting the laser spectrum, therefore loosing laser energy. Because of this, for longer pulses we could not reach the $I_{\text{IR}}^{(th)}$. Instead we maintained constant driving peak intensity ($I_{\text{IR}}$) between $1$ and $2\cross10^{14}$ W/cm$^2$. 
This was achieved by adjusting the total laser pulse energy in accordance with changes in pulse duration. As we will see later, with this intensity one can achieve about 5\% ionization degree in Ar, with which the generating conditions are below  the saturation of the harmonic yield. 
All other experimental parameters, such as focusing geometry, gas pressure and target, and interaction length, were held fixed throughout the measurement series to isolate the influence of pulse duration.
\begin{figure}[htbp]
    \centering
    \includegraphics[width=0.48\textwidth]{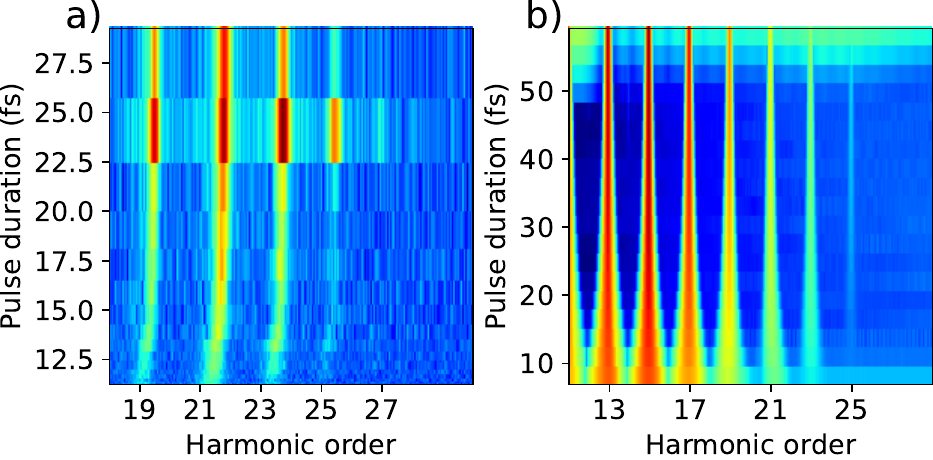}
    \caption{Measured (a)  and simulated (b) normalized harmonic spectra as the function of the driving laser pulse duration, both in logarithmic scale}
    \label{fig:constant_intensity_measured_2d}
\end{figure}
This method allows for a direct comparison of high-harmonic yields at constant intensity but varying pulse duration, which in turn affects the temporal coherence and phase-matching conditions in the medium. The experimentally obtained XUV spectra under these conditions, along with corresponding results from numerical simulations performed under equivalent macroscopic propagation conditions, are presented in Fig.~\ref{fig:constant_intensity_measured_2d}. These simulations take into account the effects of phase mismatch, absorption, and the temporal evolution of the driving field. For each pulse duration the harmonic spectrum is normalized by its maximum value. At 35\,fs the laser energy correction was not correct, it is visible from the higher then expected harmonic yield plus the extended cutoff with respect to the other pulse durations. Except for this, there is a good agreement between the experimental and numerical results.

The total high-harmonic yield serves as a crucial metric for evaluating the overall conversion efficiency of the high-harmonic generation (HHG) process. In this study, the total yield was quantified by spectrally integrating the harmonic signal over the entire measured bandwidth for each laser pulse duration. This integrated yield reflects the net energy converted from the driving infrared laser field into extreme ultraviolet (XUV) radiation within the accessible spectral range.
\begin{figure}[htbp]
    \centering
    \includegraphics[width=0.48\textwidth]{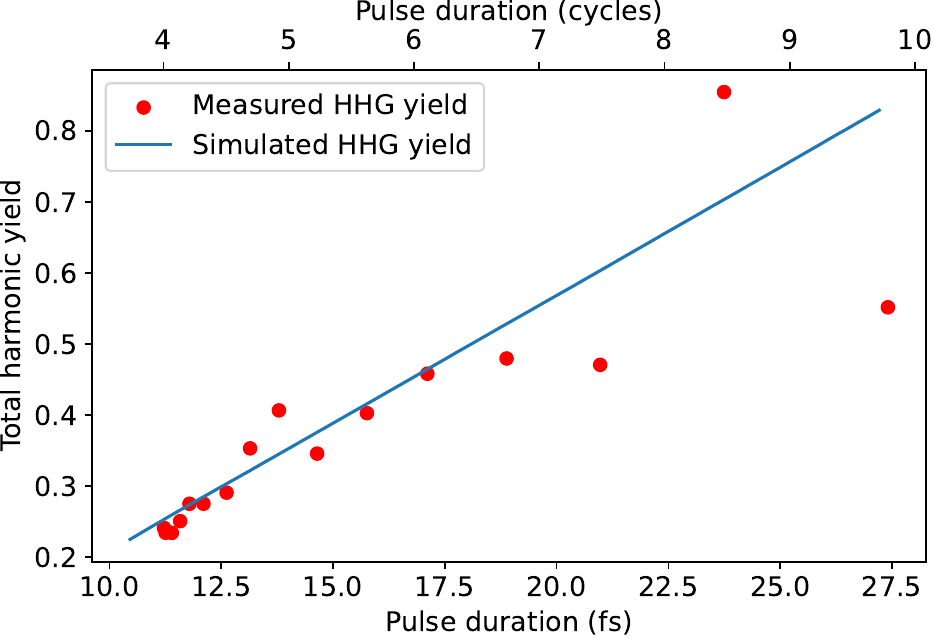}
    \caption{Dependence of the total harmonic yield on the duration of the driving IR pulse for $I_{\text{IR}}$ below the harmonic saturation intensity $I_{\text{IR}}^{(th)}$. The red dots represent the measurement and the dashed line is the simulated yield. At this ionization regime the dependence of the harmonic yield on the driving field pulse duration is linear.}
    \label{fig:total_hhg_yield_meas}
    
\end{figure}
The experimental total harmonic yield is then compared with the total yield from simulations in Fig.~\ref{fig:total_hhg_yield_meas}. The simulation exhibits a nearly linear dependence on the transform-limited pulse duration at constant intensity. The experimental data follow this simulated trend closely, indicating that the simulations capture the dominant mechanisms governing harmonic emission in this regime and validating the consistency of the experimental methodology. Therefore, from the results shown in Fig. \ref{fig:total_hhg_yield_meas} we can conclude that for $I_{\text{IR}} < I_{\text{IR}}^{(th)}$ the dependence of the harmonic yield on the driving field pulse duration is linear i.e., $E_{\text{XUV}}\propto \tau_{\text{IR}}$.


\section{The influence of the combined effect of IR pulse duration and intensity}

The findings presented in the previous section clearly demonstrate the dependence of the harmonic yield on pulse duration and support the strong agreement between the experimental results and the simulations. However, in most experiments aimed at maximizing harmonic yield, the intensity of the driving IR field is typically set just below the level that saturates the harmonic yield ($I_{\text{IR}}^{(th)}$). 
To investigate this matter, we relied on our theoretical model, which—as demonstrated in the previous section—closely agrees with the experimental observations. 

\begin{figure}[htbp]
    \centering
    \includegraphics[width=0.45\textwidth]{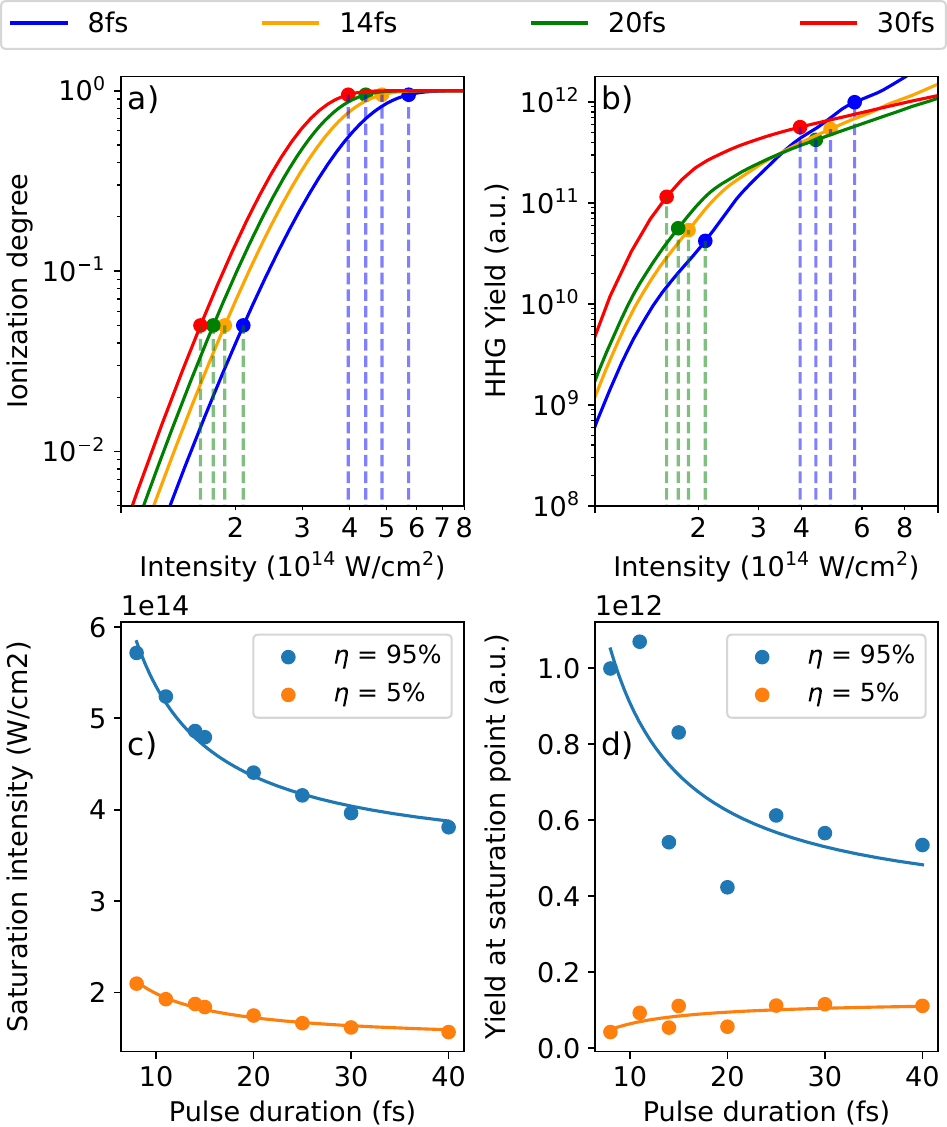}
    \caption{(a) Ionization degree and (b) simulated harmonic yield as the function of laser intensity on log-log scale for 30 fs (red), 20 fs (green), 14 fs (orange) and 8 fs (blue) laser pulse durations. Circles with blue vertical dashed line (triangles with green vertical dashed line) represent 5\% (95\%) ionization degree of the target and corresponding laser peak intensities for all the simulated pulse durations. (c) The saturation laser intensity and (d) harmonic yield at the corresponding saturation laser intensities as the function of laser pulse duration on semi-log scale for 5\% (red circles) and 95\% (green circles) ionization degrees. The dashed curves represent $\sim$ 1/$\tau$ fits.}
    \label{fig:yield_crossing}
\end{figure}

Optimizing the HHG yield requires identifying the intensity and pulse duration conditions under which the harmonic yield begins to saturate. This saturation point marks the onset of diminishing returns, where further increases in driving intensity do not lead to proportional enhancement of the harmonic signal. 
There were attempts to determine a clear $I_{\text{IR}}^{(th)}$ \cite{Weissenbilder2022}, however it relies on various assumptions. Instead of deriving an exact definition for the threshold, we demonstrate the harmonic yield dependency on the pulse duration for low (5\%) and high (95\%) levels of target ionization degrees, as it is shown in Fig. \ref{fig:yield_crossing}\,(a), where the ionization degree is plotted as the function of laser intensity for different pulse durations. It is clear that achieving the same degree of ionization with a shorter laser pulse requires higher intensity.

The corresponding harmonic yields are plotted in Fig. \ref{fig:yield_crossing}\,(b) as the function of laser intensity for both ionization degree levels. They illustrate that the saturation intensity depends strongly on the pulse duration: longer pulses tend to reach harmonic saturation at lower intensities, while shorter pulses continue to grow the yield until significantly higher intensities are reached. 
The underlying mechanism governing this behavior is primarily tunnel ionization, which can be modeled by the Ammosov–Delone–Krainov (ADK) formalism \cite{ADK1986}. As the laser intensity increases, the ionization fraction rises sharply (as shown in Fig. \ref{fig:yield_crossing}\,(a)), leading to a depletion of neutral atoms available for HHG and consequently a saturation in harmonic yield. This is shown by the two ionization levels: at 5\%, the long pulse generates more harmonics than the short ones, but it is near saturation. In contrast, the short pulses are far from saturating the yield, so increasing the laser intensity could let them produce more harmonics. This is supporting our findings in Fig. \ref{fig:total_hhg_yield_meas} where the experimental laser intensity in the focal spot was responsible for roughly $<10\%$ ionization degree. In contrast, at laser intensities corresponding to 95\% ionization degree, the long pulses have reached $I_{\text{IR}}^{(th)}$ and no longer increase the harmonic yield, but the short pulses still do.  In Fig. \ref{fig:yield_crossing}\,(c) we present how does the laser intensity, corresponding to 5\% or 95\% ionization degree, vary as the function of the laser pulse duration ($\tau$) on a semilog scale. Both cases show a $1/\tau$ trend.
Interestingly, this trend is not directly imprinted on the harmonic yield as the function of pulse duration in Fig \ref{fig:yield_crossing}\,(d). We can observe two important features. First, for laser intensities corresponding to low ionization degree, the short pulse durations produce less harmonics than long pulses, while high ionization degree maintains the $1/\tau$ tendency. Second, there are pulse durations, where the harmonic yield lies far from the observed trends, for both ionization degrees. For a rather non trivial reason it seems that the optimal harmonic yield is not always obtained for the shortest laser pulses. A similar conclusion was obtained in \cite{Westerberg2025}, through slightly different methodology.

\begin{figure}[htbp]
    \centering
    \includegraphics[width=0.48\textwidth]{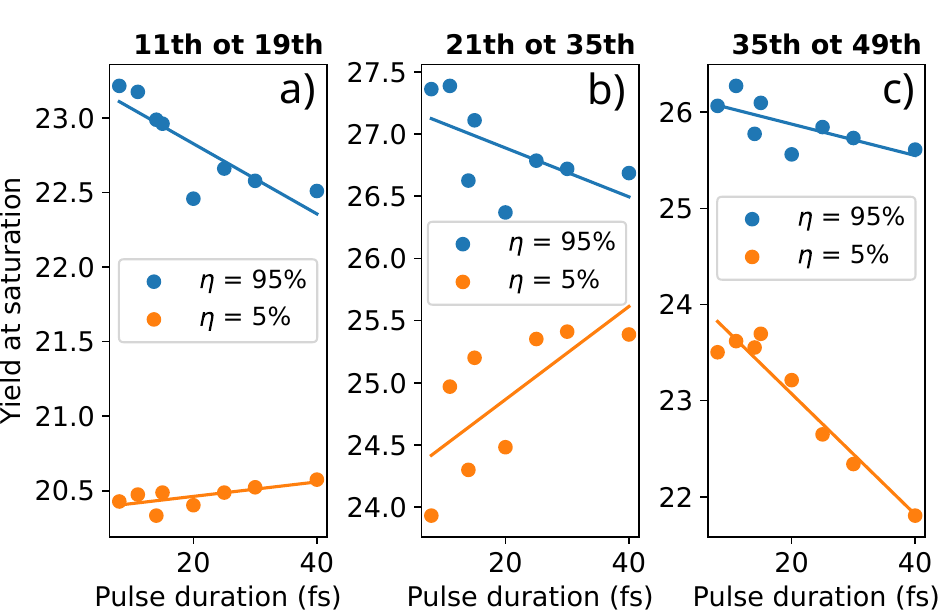}
    \caption{The harmonic yield at laser intensities corresponding to the harmonic saturation for 5\% (orange) and 95\% (blue) ionization degree as the function of laser pulse duration integrated over (a) lower-order harmonics (11th–19th), (b) plateau harmonics (21st–35th), (c) cutoff harmonics (37th–47th).}
    \label{fig:saturation_point_yield}
\end{figure}

Finally, we investigate the effect of laser pulse duration on the harmonic yield integrated over various spectral ranges in Fig. \ref{fig:saturation_point_yield} for both 5\% (orange) and 95\% (blue) ionization degrees. For the cutoff harmonics, Fig. \ref{fig:saturation_point_yield}\,c we see a drop in harmonic yields as the pulse gets longer, regardless the degree of ionization. In one hand, it is due to the fact that with shorter pulse durations, ionization occurs within a shorter time window, meaning the cutoff harmonic field—formed near the pulse peak—encounters fewer free electrons and thus experiences less dephasing. As a result, harmonic radiation can build up over a longer time, leading to higher yield \cite{Tempea2000}. On the other hand, the applied laser intensity for longer pulses is smaller, resulting in a smaller harmonic cutoff, thus having less number of photons in this spectral range. For the plateau harmonics in Fig. \ref{fig:saturation_point_yield}\,(a-b) the harmonic yield shows opposite behavior depending on the degree of the ionization. For 95\% ionization, where the longer pulses have already reached the $I_{\text{IR}}^{(th)}$ limit, substantial portion of the plateau-region harmonic emission occurs in the presence of a high free electron density. This elevated electron density destroys phase matching, thereby limiting the efficiency of harmonic generation within this spectral range. In contrast, at 5\% ionization, neither short nor long pulses produce a significant free electron population. Consequently, phase matching is better preserved over multiple optical cycles, enabling longer pulses to accumulate harmonic emission more effectively and thus achieve higher yields.

\section{Conclusion}

We have presented a systematic experimental investigation into how the HHG yield depends on the transform limited duration of the driving laser pulse — a challenging endeavor, particularly in the sub-20\,fs regime . Our work fills a critical gap, as previous studies generally lacked comprehensive experimental validation across this ultra-short pulse range.

Our measurements in Argon reveal that total harmonic yield strongly related to the pulse duration and depending on the ionization degree this relationship can vary. Below harmonic yield saturation (5\% ionization), longer pulses yield higher total XUV photon output, benefiting from extended phase-matched interaction over multiple optical cycles. Conversely, when the laser intensity reaches levels approaching 95\% ionization — shorter pulses outperform, as they avoid the detrimental effects of high free-electron density, preserve phase matching, and sustain efficient plateau and cutoff harmonic generation. 

Importantly, under conditions where harmonic yield saturation can be systematically achieved —by balancing drive intensity and gas conditions — shorter pulses emerge as the optimal choice for maximizing HHG yield. These findings provide fundamental proof and understanding on what harmonic yield should be expected when choosing between lasers of different pulse durations.

\bibliography{references}

\end{document}